\documentclass[12pt]{article}
\usepackage{amsfonts}
\usepackage{amsmath}
\usepackage{amssymb}
\usepackage{graphicx}
\usepackage{color}
\usepackage[all, knot]{xy}
\usepackage{tikz}
\usepackage{array}

\usepackage{ulem}

\usepackage[utf8]{inputenc}
\usepackage{epstopdf}
\usepackage[footnotesize]{caption}
\usepackage{amsthm}
\usepackage{enumitem}
\usepackage{mathrsfs}

\usepackage[margin=3cm]{geometry}

\def \be {\begin{equation}}
\def \ee {\end{equation}}
\def \bea {\begin{eqnarray}}
\def \eea {\end{eqnarray}}
\def \nn {\nonumber}

\def \rr {\raise.35ex\hbox{\small $\prime$}\kern-.17em{\mbox{\large $\imath$}}}

\def \dels {\partial\kern-.6em /\kern.1em}
\def \As {{A\kern-.5em / \kern.5em}}
\def \Ds {D\kern-.7em / \kern.5em}

\def \ks {k\kern-.5em /}
\def \ls {l\kern-.5em /}

\newcommand{\ci}[1]{}

\newcommand{\ba}{\begin{eqnarray}}
\newcommand{\ea}{\end{eqnarray}}
\newcommand{\bal}{\begin{align}}
\newcommand{\eal}{\end{align}}
\newcommand{\bay}[1]{\left(\begin{array}{#1}}
\newcommand{\eay}{\end{array}\right)}

\newcommand{\hide}[1]{}

\newlist{axioms}{enumerate}{2}
\setlist[axioms,1]{label=\textbf{A\arabic{axiomsi}.}, ref=A\arabic{axiomsi}}
\setlist[axioms,2]{label=\textbf{A\arabic{axiomsi}\rlap{\myEnumCounter{axiomsii}}.},%
                   ref=A\arabic{axiomsi}\myEnumCounter{axiomsii},%
                   align=parleft,%
                   leftmargin=0em,%
                   itemsep=1.4ex,%
                   before={\stepcounter{axiomsi}}}

  \usetikzlibrary{decorations.markings}

\begin{document}
\begin{titlepage}

\begin{center}

\textbf{\LARGE
Adaptive Perturbation Method in\\
 Quantum Mechanics
\vskip.3cm
}
\vskip .5in
{\large
Chen-Te Ma$^{a, b, c, d}$ \footnote{e-mail address: yefgst@gmail.com}
\\
\vskip 1mm
}
{\sl
$^a$
Guangdong Provincial Key Laboratory of Nuclear Science, 
\\
Institute of Quantum Matter,
South China Normal University, Guangzhou 510006, 
Guangdong, China.
\\
$^b$
School of Physics and Telecommunication Engineering,
\\
 South China Normal University, 
 Guangzhou 510006, Guangdong, China.
\\
$^c$
Guangdong-Hong Kong Joint Laboratory of Quantum Matter,
\\
 Southern Nuclear Science Computing Center, 
South China Normal University,
Guangzhou 510006, Guangdong, China.
\\
$^d$
The Laboratory for Quantum Gravity and Strings,
\\
 Department of Mathematics and Applied Mathematics,
University of Cape Town,
Private Bag, Rondebosch 7700, South Africa.
}
\\
\vskip 1mm
\vspace{40pt}
\end{center}
\begin{abstract}
The adaptive perturbation chooses a non-standard decomposition. 
The Hamiltonian becomes a sum of solvable and perturbation parts.
We calculate the spectrum using the adaptive perturbation method at the leading-order to compare to numerical solutions.
The maximum deviation is around $5\%$ for different coupling regions. 
A perturbation study relies on whether a choice of leading-order is suitable. 
Our result with different parameters should show that the adaptive perturbation method provides appropriate saddle points to all coupling regions.
In the end, we show that the perturbation parameters should not be a coupling constant.
\end{abstract}
\end{titlepage}

\section{Introduction}
\label{sec:1}
\noindent
The spectrum of black-body radiation does not have a precise match from classical physics. 
The phenomenon cannot have an interpretation without knowing about quantized energy. 
This observation first introduced a discrete energy level in a physical system and established quantum mechanics for the atomic scale.
\\

\noindent
Because the spectrum is only solvable in a few systems, people use a perturbation method to study quantum physics in a weakly interacting system. 
We begin with a solvable non-interacting theory and study the effect of interaction perturbatively.
This procedure is standard. 
We can label quantum states of oscillator systems without any interaction by particle numbers (Fock state). 
Since the coupling is weak, one can do an expansion by the Fock state. 
The perturbation method is systematic, but it is hard to apply to a strongly coupled system \cite{Weisskopf:1930au, Feynman:1963uxa, Ma:2018efs}. 
Even for a weak coupling expansion in 0-dimensional $\lambda x^4$ theory, the series is only asymptotically convergent. 
To obtain complete information about a physical system, improving the perturbation method is necessary. 
\\

\noindent
People considered a different unperturbed Hamiltonian to obtain a convergent result \cite{Halliday:1979vn, Halliday:1979xh}. 
The adaptive perturbation method also considered a non-conventional unperturbed Hamiltonian \cite{Weinstein:2005kw, Weinstein:2005kx}. 
The total Hamiltonian is the sum of solvable and perturbation parts \cite{Weinstein:2005kw, Weinstein:2005kx}. 
The diagonal elements of a Fock space are the unperturbed part. 
The elements include an interacting sector at the leading order. One also introduced a variable without changing the canonical relation \cite{Weinstein:2005kw, Weinstein:2005kx}. 
We determine this variable by the minimized expectation value of the energy with a Fock state \cite{Weinstein:2005kw, Weinstein:2005kx}. The solvable part is different from the non-interacting sector in general. The above is the main trick of this method. The central question that we would like to address in this letter is the following: {\it How well studied for the adaptive perturbation method?} To answer this question, we analyze the saddle points from different parameters \cite{Curcio:2018}. 
\\

\noindent
In this letter, we study a model from an exact solution of the solvable part. 
We then analyze the deviation of the spectrum between the solvable part and the numerical solution for different parameters. 
Our result provides a {\it quantitative study} by comparing the analytical solutions to {\it numerical solutions} for {\it all coupling regions}. 
We then show that the {\it perturbation parmeter} should {\it not} be a {\it coupling constant}. 
Our perturbation study of the spectrum is also {\it fully analytical} in all coupling regions. 
The system cannot have such an analytical study before. 

\section{Adaptive Perturbation Method in Quantum Mechanics}
\label{sec:2}
\noindent
We first demonstrate how to determine the leading-order term from the Hamiltonian (corresponds to $\omega^2=0$, where $\omega$ is the frequency)
\bea
H_1=\frac{p^2}{2}+\lambda_1 \frac{x^4}{6}+\lambda_2\frac{x^6}{120},
\eea
 where $p$ and $x$ are the momentum and position operators, and $\lambda_1$ and $\lambda_2$ are the coupling constants. The $p$ and $x$ satisfy the usual commutation relation $\lbrack p, x\rbrack=-i$. Now we introduce the $A_{\gamma}^{\dagger}$ and $A_{\gamma}$ as that \cite{Weinstein:2005kw, Weinstein:2005kx}: 
$x=(A_{\gamma}^{\dagger}+A_{\gamma})/\sqrt{2\gamma}$ and $p=i\sqrt{\gamma/2}(A_{\gamma}^{\dagger}-A_{\gamma})$.
The commutation relation between $A_{\gamma}$ and $A_{\gamma}^{\dagger}$ is 
$\lbrack A_{\gamma}, A_{\gamma}^{\dagger}\rbrack=1$. 
The additional variable $\gamma$ does not modify the commutation relation. 
The operators acting on a quantum state gives that \cite{Weinstein:2005kw, Weinstein:2005kx}: $N_{\gamma}|n_{\gamma}\rangle=n|n_{\gamma}\rangle$ and 
$A_{\gamma}|0_{\gamma}\rangle=0$, where $N_{\gamma}\equiv A_{\gamma}^{\dagger}A_{\gamma}$.
\\

\noindent
We decompose the Hamiltonian to a solvable part and a perturbation part \cite{Weinstein:2005kw, Weinstein:2005kx}. 
The diagonal elements of a Fock space are the solvable part \cite{Weinstein:2005kw, Weinstein:2005kx}. 
In other words, the solvable part of the Hamiltonian $H_0(\gamma)$ can be written in terms of the $N_{\gamma}$
\bea
&&H_0(\gamma)
\nn\\
&=&\frac{\gamma}{4}(2N_{\gamma}+1)+\frac{\lambda_1}{4\gamma^2}\bigg(N_{\gamma}^2+N_{\gamma}+\frac{1}{2}\bigg)
\nn\\
&&
+\frac{\lambda_2}{4\gamma^3}\bigg(\frac{1}{12}N_{\gamma}^3
+\frac{29}{240}N_{\gamma}^2+\frac{1}{6}N_{\gamma}+\frac{1}{16}\bigg).
\nn\\
\eea
The expectation value of the energy is:
\bea
&&E_n(\gamma)
\nn\\
&\equiv&\langle n_{\gamma}| H_1(\gamma)|n_{\gamma}\rangle
\nn\\
&=&\langle n_{\gamma}|H_0(\gamma)|n_{\gamma}\rangle
\nn\\
&=&
\frac{\gamma}{4}(2n+1)
\nn\\
&&+\frac{\lambda_1}{4\gamma^2}\bigg(n^2+n+\frac{1}{2}\bigg)
\nn\\
&&
+\frac{\lambda_2}{4\gamma^3}\bigg(\frac{1}{12}n^3+\frac{29}{240}n^2+\frac{1}{6}n+\frac{1}{16}\bigg).
\nn\\
\eea
We still have an undetermined variable $\gamma$. 
To fix this variable, we choose the minimized expectation value of the energy to determine the value \cite{Weinstein:2005kw, Weinstein:2005kx}. 
The minimized energy occurs when $\gamma>0$ satisfies that
\bea
&&
\gamma^4
-2\lambda_1\frac{n^2+n+\frac{1}{2}}{2n+1}\gamma
\nn\\
&&
-3\lambda_2\frac{\frac{1}{12}n^3+\frac{29}{240}n^2+\frac{1}{6}n+\frac{1}{16}}{2n+1}=0.
\eea
We then choose the minimized expectation value of the energy as the unperturbed spectrum. 
Because the form of the exact solution is ugly, we do not write it in this letter.
\\

\noindent
We use the naive discretization for the kinematic term 
\bea
p^2\psi\rightarrow-\frac{\psi_{j+1}-2\psi_j+\psi_{j-1}}{a^2}.
\eea
The $\psi_j$ is the eigenfunction for the lattice theory, and $a$ is the lattice spacing. 
We label the lattice index by $j=1, 2, \cdots, n_l$, where $n_l$ is the number of lattice points.  
In the numerical solution, we choose the lattice size and the number of lattice points:
\bea
L=8\equiv \frac{n_la}{2};\qquad n_l=1024.
\eea
\\

\noindent
We first turn off the $\lambda_2$. We then observe that the $E_n(\gamma)_{\mathrm{min}}$ deviates the numerical result within $1\%$ when $n>1$ and $\lambda_1=16$ in Table \ref{c116}. For $n\le 2$, the maximum deviation is around $2\%$. 
Hence the adaptive perturbation method at the leading-order already shows a quantitative result in the strongly coupled region. 
We also observe a similar deviation for $\lambda_1=0.25$ in Table \ref{c1025}. 
Therefore, the adaptive perturbation method should be suitable for all coupling regions without any scarifying. 
We then turn on the $\lambda_2$ and in Tables \ref{c216256} and \ref{c20254}. 
The behavior of deviations is also similar (the maximum deviation is around $5\%$).
For a high quantum number, it becomes harder to obtain an accurate result from a numerical study. 
Hence this result shows that the adaptive perturbation method is quite helpful for such a region. 
We define the deviation as $\bigg(100\times\big|\big((\mathrm{Numerical\ Solution})- E_n(\gamma)_{\mathrm{min}}\big)\big/(\mathrm{Numerical\ Solution})\big|\bigg)\%$ in all Tables.
\\

\begin{table}[!htb]
\centering
\begin{tabular}{ |m{1em} | m{2cm}| m{2cm}|m{2cm}| } 
\hline
\textbf{$n$} & \textbf{$E_n(\gamma)_{\mathrm{min}}$} &\textbf{Numerical Solution}&\textbf{Deviation} \\ 
\hline
$0$ & 0.944& 0.926&1.943\%\\ 
\hline
$1$ & 3.361& 3.319&1.265\%\\ 
\hline
$2$ & 6.496& 6.512&0.245\%\\ 
\hline
$3$ & 10.11& 10.17&0.589\%\\ 
\hline
$4$ & 14.098& 14.201&0.725\%\\ 
\hline
$5$ & 18.398& 18.545&0.792\%\\ 
\hline
$6$ & 22.97& 23.162&0.828\%\\ 
\hline
$7$ & 27.785& 28.022&0.845\%\\ 
\hline
\end{tabular}
\caption{The comparison between the $E_n(\gamma)_{\mathrm{min}}$ and the numerical solutions for the $\lambda_1=16$, $\lambda_2=0$, and $\omega^2=0$.}
\label{c116}
\end{table}

\begin{table}[!htb]
\centering
\begin{tabular}{ |m{1em} | m{2cm}| m{2cm}|m{2cm}| } 
\hline
\textbf{$n$} & \textbf{$E_n(\gamma)_{\mathrm{min}}$} &\textbf{Numerical Solution}&\textbf{Deviation} \\ 
\hline
$0$ & 0.236& 0.231&2.164\%\\ 
\hline
$1$ & 0.84& 0.829&1.326\%\\ 
\hline
$2$ & 1.624& 1.628&0.245\%\\ 
\hline
$3$ & 2.527& 2.543&0.629\%\\ 
\hline
$4$ & 3.524& 3.551&0.76\%\\ 
\hline
$5$ & 4.599& 4.637&0.819\%\\ 
\hline
$6$ & 5.742& 5.792&0.863\%\\ 
\hline
$7$ & 6.946& 7.009&0.898\%\\ 
\hline
\end{tabular}
\caption{The comparison between the $E_n(\gamma)_{\mathrm{min}}$ and the numerical solutions for the $\lambda_1=0.25$, $\lambda_2=0$, and $\omega^2=0$.}
\label{c1025}
\end{table}

\begin{table}[!htb]
\centering
\begin{tabular}{ |m{1em} | m{2cm}| m{2cm}|m{2cm}| } 
\hline
\textbf{$n$} & \textbf{$E_n(\gamma)_{\mathrm{min}}$} &\textbf{Numerical Solution}&\textbf{Deviation} \\ 
\hline
$0$ & 1.117& 1.075&3.906\%\\ 
\hline
$1$ & 4.047& 3.949&2.481\%\\ 
\hline
$2$ & 7.993& 7.989&0.05\%\\ 
\hline
$3$ & 12.724& 12.831&0.833\%\\ 
\hline
$4$ & 18.109& 18.338&1.248\%\\ 
\hline
$5$ & 24.067& 24.426&1.469\%\\ 
\hline
$6$ & 30.54& 31.038&1.604\%\\ 
\hline
$7$ & 37.486& 38.13&1.688\%\\ 
\hline
\end{tabular}
\caption{The comparison between the $E_n(\gamma)_{\mathrm{min}}$ and the numerical solutions for the $\lambda_1=16$, $\lambda_2=256$, and $\omega^2=0$.}
\label{c216256}
\end{table}

\begin{table}[!htb]
\centering
\begin{tabular}{ |m{1em} | m{2cm}| m{2cm}| m{2cm}| } 
\hline
\textbf{$n$} & \textbf{$E_n(\gamma)_{\mathrm{min}}$}&\textbf{Numerical Solution} &\textbf{Deviation} \\ 
\hline
$0$ & 0.343& 0.326&5.214\%\\ 
\hline
$1$ & 1.258& 1.218&3.284\%\\ 
\hline
$2$ & 2.512& 2.507&0.199\%\\ 
\hline
$3$ & 4.039& 4.079&0.98\%\\ 
\hline
$4$ & 5.795& 5.884&1.512\%\\ 
\hline
$5$ & 7.753& 7.894&1.786\%\\ 
\hline
$6$ & 9.892& 10.089&1.952\%\\ 
\hline
$7$ & 12.197& 12.454&2.063\%\\ 
\hline
\end{tabular}
\caption{The comparison between the $E_n(\gamma)_{\mathrm{min}}$ and the numerical solutions for the $\lambda_1=0.25$, $\lambda_2=4$, and $\omega^2=0$.}
\label{c20254}
\end{table}

\section{Mass Term}
\label{sec:3}
\noindent
Now we introduce a mass term to investigate the deviations. 
For a non-vanishing mass term, the small quantum number and coupling constant case should close to the harmonic oscillator. Therefore, we expect that the non-vanishing mass term can play a suitable infrared cut-off. The Hamiltonian becomes 
\bea
H_2=\frac{p^2}{2}+\omega^2\frac{x^2}{2}+\lambda_1\frac{x^4}{6}+\lambda_2\frac{x^6}{120}.
\eea
The expectation value of the energy is 
\bea
&&E_n(\gamma)
\nn\\
&=&\frac{\gamma}{4}(2n+1)
+\frac{\omega^2}{4\gamma}(2n+1)
\nn\\
&&
+\frac{\lambda_1}{4\gamma^2}\bigg(n^2+n+\frac{1}{2}\bigg)
\nn\\
&&
+\frac{\lambda_2}{4\gamma^3}\bigg(\frac{1}{12}n^3+\frac{29}{240}n^2+\frac{1}{6}n+\frac{1}{16}\bigg).
\eea
The minimized expectation value of the energy occurs when the $\gamma>0$ satisfies that
\bea
&&
\gamma^4-\omega^2\gamma^2-\lambda_1\frac{2n^2+2n+1}{2n+1}\gamma
\nn\\
&&
-\frac{\lambda_2}{80}\frac{20n^3+29n^2+40n+15}{2n+1}
=0,
\eea
\\

\noindent
We first turn off the $\lambda_2$. 
Table \ref{c316} exhibits a quantitative comparison (the maximum deviation is around $1.5\%$). 
We also observed similar behavior (the maximum deviation is about $3\%$)  by turning on the $\lambda_2$ in Table \ref{c416256}. 
 From the above Tables, we observe that introducing the mass term does not lose a quantitative study. In the above Tables, we choose $\omega=1$.
\\

\begin{table}[!htb]
\centering
\begin{tabular}{ |m{1em} | m{2cm}| m{2cm}| m{2cm}| } 
\hline
\textbf{$n$} & \textbf{$E_n(\gamma)_{\mathrm{min}}$} &\textbf{Numerical Solution}&\textbf{Deviation} \\ 
\hline
$0$ & 1.041& 1.026&1.461\%\\ 
\hline
$1$ & 3.607& 3.571&1.008\%\\ 
\hline
$2$ & 6.852& 6.863&0.16\%\\ 
\hline
$3$ & 10.56& 10.611&0.48\%\\ 
\hline
$4$ & 14.631& 14.723&0.624\%\\ 
\hline
$5$ & 19.009& 19.143&0.699\%\\ 
\hline
$6$ & 23.655& 23.83&0.734\%\\ 
\hline
$7$ & 28.539& 28.7581&0.761\%\\ 
\hline
\end{tabular}
\caption{The comparison between the $E_n(\gamma)_{\mathrm{min}}$ and the numerical solutions for the $\lambda_1=16$, $\lambda_2=0$, and $\omega^2=1$.}
\label{c316}
\end{table}

\begin{table}[!htb]
\centering
\begin{tabular}{ |m{1em} | m{2cm}| m{2cm}| m{2cm}| } 
\hline
\textbf{$n$} & \textbf{$E_n(\gamma)_{\mathrm{min}}$} &\textbf{Numerical Solution}&\textbf{Deviation}  \\ 
\hline
$0$ & 1.195& 1.159&3.106\%\\ 
\hline
$1$ & 4.242& 4.154&2.118\%\\ 
\hline
$2$ & 8.266& 8.26&0.072\%\\ 
\hline
$3$ & 13.059& 13.157&0.744\%\\ 
\hline
$4$ & 18.498& 18.712&1.143\%\\ 
\hline
$5$ & 24.503& 24.844&1.372\%\\ 
\hline
$6$ & 31.019& 31.496&1.514\%\\ 
\hline
$7$ & 38.005& 38.625&1.605\%\\ 
\hline
\end{tabular}
\caption{The comparison between the $E_n(\gamma)_{\mathrm{min}}$ and the numerical solutions for the $\lambda_1=16$, $\lambda_2=256$, and $\omega^2=1$.}
\label{c416256}
\end{table}

\begin{figure}[!htb]
\includegraphics[width=0.5\textwidth]{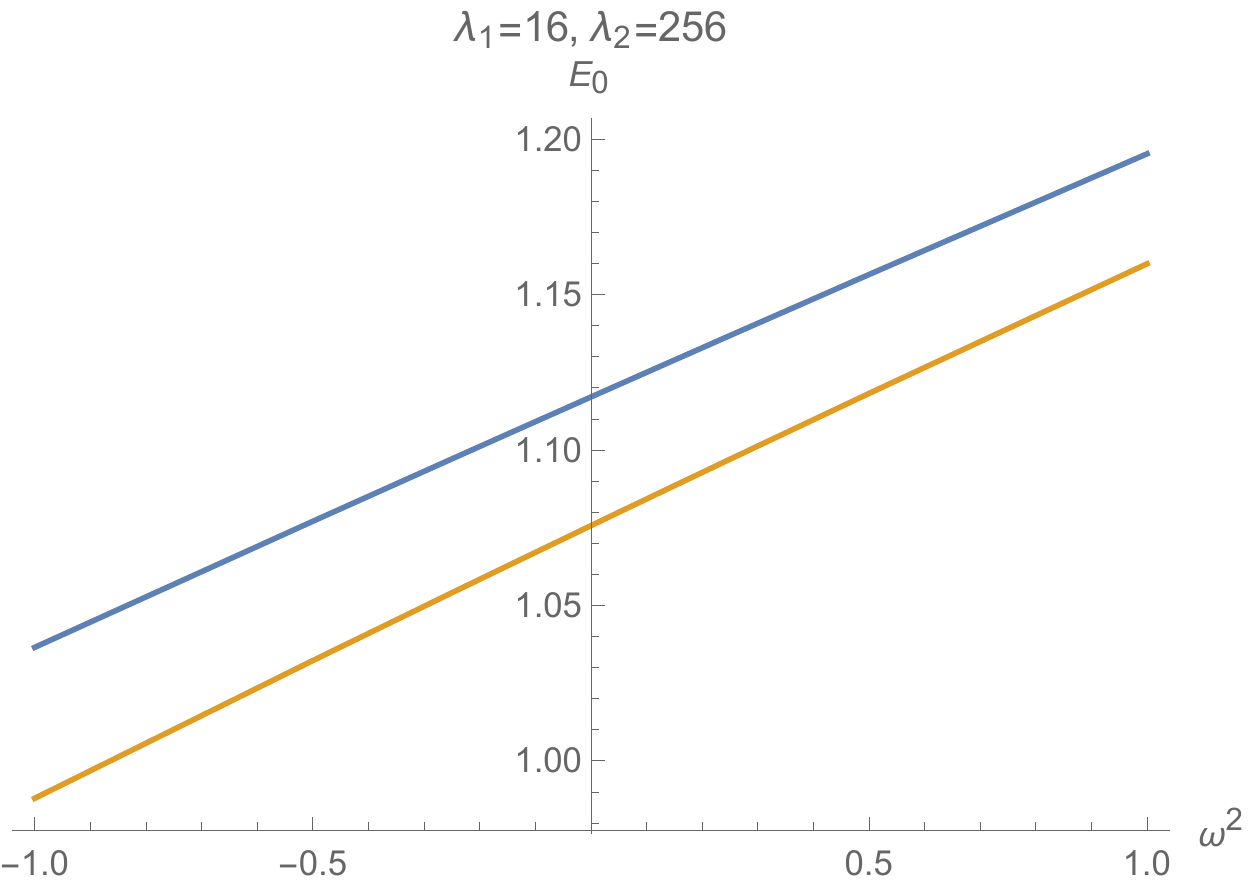}
\includegraphics[width=0.5\textwidth]{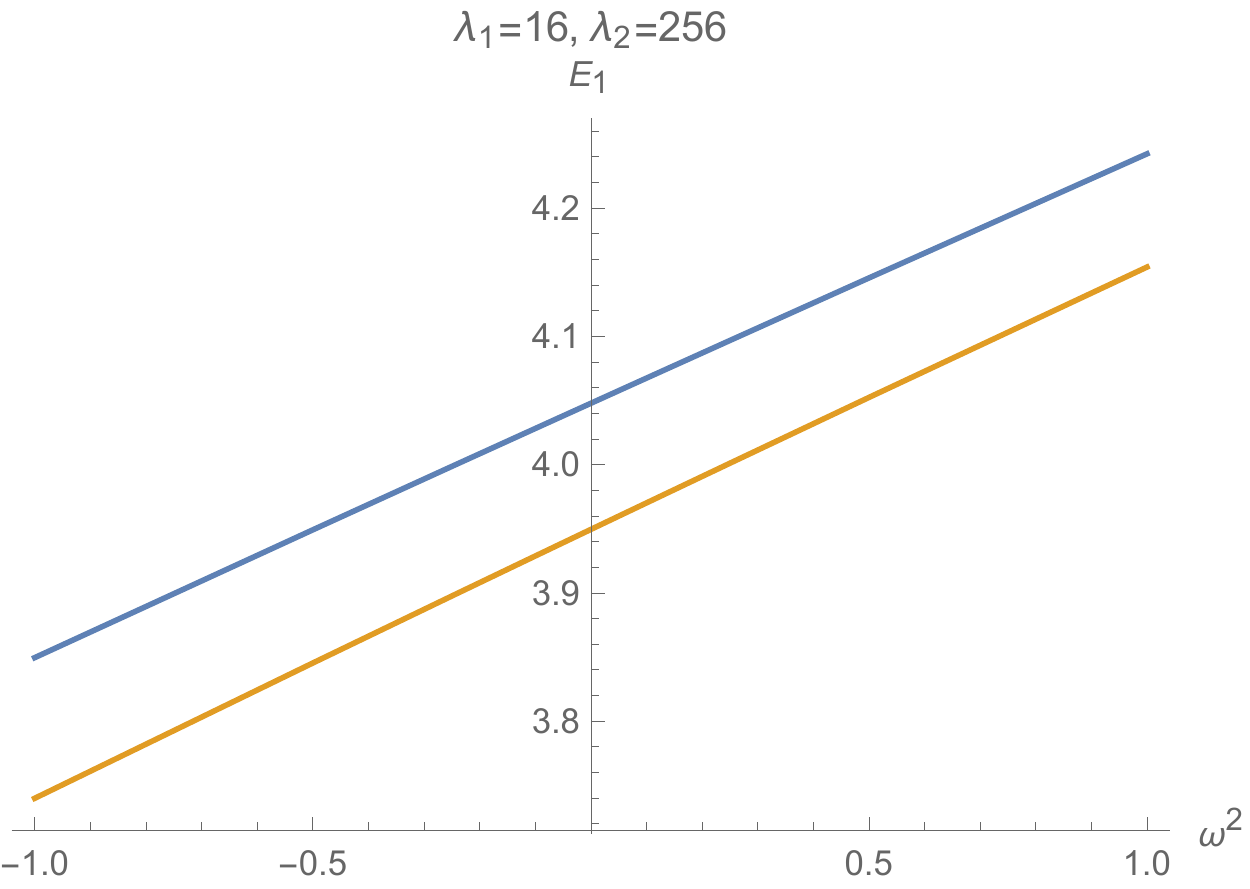}
\caption{We compare the result of adaptive perturbation to the numerical solution for different $\omega^2$. 
The $E_0$ and $E_1$ are the ground state and first-excited energies, respectively. } 
\label{comparison}
\end{figure}

\noindent 
Now we switch the sign of the mass term. 
The appearance of an inverted mass term ($\omega^2<0$) provides the degenerate vacuum state. 
In Quantum Field Theory, this case corresponds to spontaneous symmetry breaking (SSB). 
Without an interacting term, an inverted oscillator does not have a bound state. 
Therefore, the SSB is necessary when people obtain a perturbation or an observed result. 
Hence the study of this case is unavoidable. 
We show the result of $\omega^2=-1$ in Table \ref{c-1}. 
\begin{table}[!htb]
\centering
\begin{tabular}{ |m{1em} | m{2cm}| m{2cm}| m{2cm}| } 
\hline
\textbf{$n$} & \textbf{$E_n(\gamma)_{\mathrm{min}}$} &\textbf{Numerical Solution}&\textbf{Deviation}  \\ 
\hline
$0$ & 1.036& 0.987&4.964\%\\ 
\hline
$1$ & 3.849& 3.739&2.941\%\\ 
\hline
$2$ & 7.716& 7.716&0\%\\ 
\hline
$3$ & 12.385& 12.502&0.935\%\\ 
\hline
$4$ & 17.717& 17.961&1.358\%\\ 
\hline
$5$ & 23.628& 24.006&1.574\%\\ 
\hline
$6$ & 30.058& 30.578&1.7\%\\ 
\hline
$7$ & 36.965& 37.634&1.777\%\\ 
\hline
\end{tabular}
\caption{The comparison between the $E_n(\gamma)_{\mathrm{min}}$ and the numerical solutions for the $\lambda_1=16$, $\lambda_2=256$, and $\omega^2=-1$.}
\label{c-1}
\end{table}
The maximum deviation is around $5\%$ for the strong coupling. 
Hence we conclude that adaptive perturbation method can be applied to the inverted anharmonic oscillator.
The result is helpful before an implementation to Quantum Field Theory. 
In the end, we compare the result of adaptive perturbation to the numerical solution for different $\omega^2$ in Fig. \ref{comparison}. 
We observe that the difference of perturbation and numerical solutions does not have a clear dependence on $\omega^2$. 
When one considers $\omega^2\le 0$, using the conventional method will show non-physical results. 
The adaptive perturbation method still gives quantitative and physical results. 

\section{Time-Independent Perturbation}
\label{sec:4}
\noindent
We study the time-independent perturbation of the $H_1$ with $\lambda_2=0$ \cite{Weinstein:2005kw, Weinstein:2005kx}. 
The result implies that a coupling constant is not a perturbation parameter. The eigenenergy calculated by the time-independent perturbation is 
\bea
&&E_n
\nn\\
&=&
E_n^{(0)}+\langle n^{(0)}|V|n^{(0)}\rangle+\sum_{k\neq n}
\frac{|\langle k^{(0)}|V|n^{(0)}\rangle|^2}{E_n^{(0)}-E_{k,n}^{(0)}}+\cdots,
\nn\\
\eea
The $E_n^{0}$ is the $n$-th unperturbed eigenenergy. 
The $|n^{(0)}\rangle$ is the $n$-th unperturbed eigenstate. 
The $V$ is a perturbation as that $H_1\equiv H_0+V$, and the $E_n^{(0)}$ is $E_n(\gamma)_{\mathrm{min}}$. The $E_{k, n}$ is the $k$-th unperturbed eigenenergy with the $n$-th $\gamma$. We then show that each term is at the same order of the coupling constant $\lambda_1^{1/3}$. This proof is not a bad result and gives consistency to the spectrum because we can apply the transformations, $x\rightarrow x/\lambda_1^{1/6}$ and $p\rightarrow\lambda_1^{1/6}p$, to show that $H_1\propto\lambda_1^{1/3}$. In the end, we also find that $|\langle k^{(0)}|V|n^{(0)}\rangle|^2$ contributes $n^2$ and $E_n^{(0)}-E_{k, n}^{(0)}$ also contributes so when a quantum number is large enough. Hence no divergence comes from a summation of the quantum numbers. 
Indeed, it is also due to using $E_n(\gamma)_{\mathrm{min}}$ because it includes information about the coupling constant. 
Hence it is why the saddle-point can show a quantitative result. 

\section{Outlook}
\label{sec:5}
\noindent
We analyzed the deviation between $E_n(\gamma)_{\mathrm{min}}$ and a numerical solution from different parameters. Because $E_n(\gamma)_{\mathrm{min}}$ is the leading-order result \cite{Weinstein:2005kw, Weinstein:2005kx}, how the perturbation works well relies on a suitable choice of a saddle-point. We also showed that the perturbation parameter is not a coupling constant from the explicit Hamiltonian $H_1$ with the $\lambda_2=0$. 
Our perturbation study is analytical in all coupling regions. 
It is non-trivial because various perturbation studies lose a fully analytical way in a strongly coupled region. 
Here we only demonstrated the leading-order result. 
However, the leading order already implies that all perturbation terms should also have an exact solution. 
\\

\noindent
One non-trivial application of quantum mechanics is to observe whether a spectrum can have a universal rule when a phase transition occurs. 
This application can teach us how to probe the problems of critical points in Quantum Chromodynamics (QCD). 
The QCD does not have an exact solution to the above problem because it is a non-integrable model. 
One suitable and practical way is to use a suitable saddle-point to do a perturbation without doing a very high-order calculation for obtaining physical information. 
We show that the adaptive perturbation method gives such a study to Quantum Mechanics.
The perturbation problem and theoretical formulation in harmonic oscillator should be similar to scalar field theory. 
Extending our study to scalar fields should be the first step for studying QCD in a strongly coupled region.

\section*{Acknowledgments}
\noindent
The author would like to thank Jiunn-Wei Chen, Bo Feng, Pak Hang Chris Lau, Gang Yang, and Ellis Ye Yuan for their discussion.
\\

\noindent
The author acknowledges the Post-Doctoral International Exchange Program; 
China Postdoctoral Science Foundation, Postdoctoral General Funding: Second Class (Grant No. 2019M652926); 
Foreign Young Talents Program (Grant No. QN20200230017). 
The author also would like to thank Nan-Peng Ma for his encouragement.
\\

\noindent
The author would like to thank the National Center for Theoretical Sciences at the National Tsing Hua University, Sun Yat-Sen University, and Zhejiang Institute of Modern Physics at the Zhejiang University.
\\

\noindent
Discussions during the workshops, ``East Asia Joint Workshop on Fields and Strings 2019 and 12th Taiwan String Theory Workshop'' and ``Composite 2019: Hunting Physics in Higgs, Dark Matter, Neutrinos, Composite Dynamics and Extra-Dimensions'', were helpful to complete this work. 


\end{document}